\pdfoutput=1
\documentclass[preprint,amsfonts,amsmath,amssymb,aps,physrev,titlepage,nofootinbib]{revtex4-2}

\usepackage[margin=1in]{geometry}
\usepackage{graphicx} 
\usepackage[utf8]{inputenc}
\usepackage[T1]{fontenc} 
\usepackage{microtype} 
\usepackage[scr]{rsfso} 
\usepackage{physics} 
\usepackage{tensor} 
\usepackage[svgnames]{xcolor} 
    \definecolor{accent}{HTML}{df2d16}
\usepackage{tikz}
\usetikzlibrary{arrows.meta, calc, decorations.pathmorphing}
\usepackage[allcolors=blue,colorlinks]{hyperref} 
\usepackage{ORCIDinREVTeX} 

\begin{document}

\title{The Measure of a Mass}

\author{Níckolas de \surname{Aguiar Alves}}
\orcid{0000-0002-0309-735X}
\email{alves.nickolas@ufabc.edu.br}    
\affiliation{Center for Natural and Human Sciences, Federal University of ABC,\\Av. dos Estados, 5001, 09210-580, Santo André, São Paulo, Brazil}

\author{Bruno \surname{Arderucio Costa}}
\orcid{0000-0001-5182-2010}
\email[Corresponding author: ]{bcosta@troy.edu}
\affiliation{Center for Relativity and Cosmology, Troy University,\\ Troy, Alabama, 36082, USA}

\date{March 19, 2025}

\begin{abstract}
The concept of mass is central to any theory of gravity. Nevertheless, defining mass in general relativity is a difficult task, and even when it can be accomplished, we still need to investigate whether the typical properties of mass in Newtonian gravity are still true in Einsteinian gravity. In this essay, we discuss ``the measure of a mass'' in relativity by considering some of the many different definitions (Komar, ADM, and Bondi) and how they are related. Finally, we discuss when and whether the mass is positive, as is usually expected, and which physical properties of matter and gravity can ensure this result.

\vspace{24pt}
\noindent{}{Essay written for the \href{https://www.gravityresearchfoundation.org/competition}{Gravity Research Foundation 2025 Awards for Essays on Gravitation}.}
\end{abstract}

\maketitle
\thispagestyle{empty}

\pagenumbering{arabic}
It is said that the first great episode of \emph{Star Trek: The Next Generation} is Season 2's \emph{The Measure of a Man}. The episode discusses to what extent Lt. Cdr. Data, an android, can be considered a sentient being and assigned the rights one fully deserves. If an android is merely the collection of its components, it may lack consciousness. If so, would he still have the right to deny being, for example, dismantled to advance science? How do you measure whether a biological being and an android are both conscious creatures in the same way? 

Although the main problem posed by \emph{The Measure of a Man} is far more complex than we can answer with present-day physics, at a fundamental level it invites us to rethink whether the whole is more than the sum of its parts and how (and whether) we should update old concepts to fit new information. These questions permeate all of physics; a simple example of interest is the concept of mass in general relativity. Newtonian physics could, at first, tempt us to say that a macroscopic mass is nothing but an agglomerate of microscopic bunches of matter. But is this really the measure of a mass? 

In Newtonian gravity, ``mass'' refers to a measure of the ``power of attraction'' of a matter distribution. To codify the properties of a distribution in a single number, one considers a monopole moment. Given many test masses accelerating toward the central distribution with acceleration \(\vb{a}\), the central mass is given by
\begin{equation}\label{eq: active-mass-newton}
    M = - \frac{1}{4 \pi G} \oint_{\mathbb S^2} \vb{a} \vdot \dd{\vb{S}}.
\end{equation}
Here, the integration surface $\mathbb S^2$ is understood to be ``at infinity'' to enclose all matter content.

It is not straightforward to generalize Eq. \eqref{eq: active-mass-newton} to general relativity because the gravitational field and spacetime itself are now two facets of the same physical entity. Furthermore, since space and time become interlaced, ``integrating over a sphere at infinity'' may not be as objective as it seems. As a consequence, there are many notions of ``total mass'' in Einstein's gravity.

These subtleties can be easily addressed in stationary, asymptotically flat spacetimes. The first condition singles out a class of observers ``at rest,'' and their accelerations serve as a measure of how strong the gravitational pull is. The second condition ensures that these observers can be considered ``at infinity,'' where the monopole contribution dominates. These requirements may seem strong. Nevertheless, the resulting notion of mass---known as the Komar mass \cite{komar1959CovariantConservationLaws}---can later be used as a reference for more general definitions. \emph{A posteriori}, distinct notions of mass defined under less stringent conditions should agree in this regime---and they do \cite{ashtekar1979ConservedQuantitiesGeneral}.

Within a stationary setup, we expect mass to be conserved in time, for it is an expression of energy. Since a timelike Killing vector field is available, we have a notion of time-translation symmetry at hand, and the Komar mass can then be understood as a Noether charge for this symmetry. However, if we consider more general spacetimes, mass may or may not be conserved. Mathematically, this corresponds to asking whether the definition under consideration depends on the relativistic analog of the integration surface \(\mathbb S^2\) in Eq.~\eqref{eq: active-mass-newton} as it is ``dragged'' to the future.

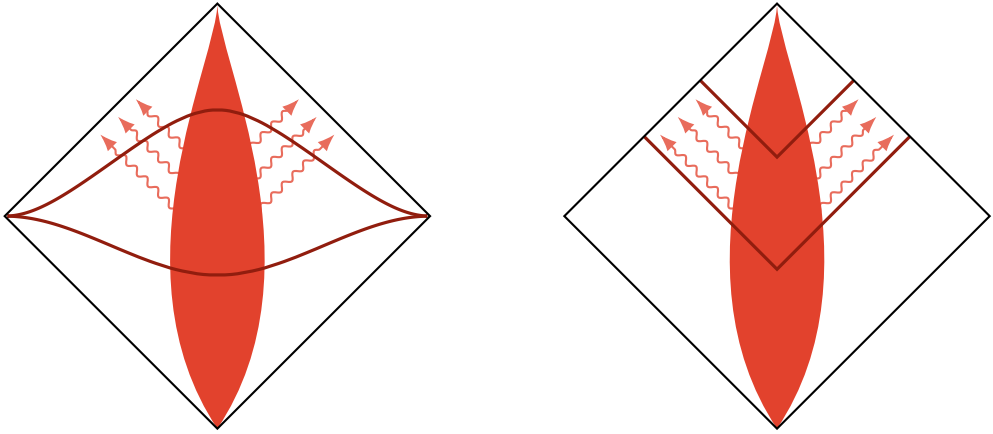
\begin{figure}
    \centering
    \null\hfill
    \begin{tikzpicture}[scale=0.9]
        \begin{scope}[decoration={snake,amplitude=.4mm,segment length=2mm,post length=2mm},>=Latex,thick,transparency group,opacity=0.7,accent]
            \foreach \t in {-30,0,30}{                
                \draw[decorate,->] ($(0,{rad(\t)})$) -- ($({(pi-rad(\t))/2},{(pi+rad(\t))/2})+(-0.1,-0.1)$);
                \draw[decorate,->] ($(0,{rad(\t)})$) -- ($({-(pi-rad(\t))/2},{(pi+rad(\t))/2})+(0.1,-0.1)$);
            };
        \end{scope}
        
        \fill[color=accent!90!white,very thick] (0,-pi) .. controls ($(pi/2,-pi/4)$) and ($(0,3*pi/4)$) .. (0,pi) .. controls ($(0,3*pi/4)$) and ($(-pi/2,-pi/4)$) .. (0,-pi);
        
        \foreach \t in {-50, 90}{                
            \draw[domain=-50:50,smooth,variable=\r,samples=400,accent!65!black,very thick] plot ({rad(atan(tan(\t/2)+\r))-rad(atan(tan(\t/2)-\r))},{rad(atan(tan(\t/2)+\r))+rad(atan(tan(\t/2)-\r))});
        };

        \draw[thick] (0,-pi) -- (pi,0) -- (0,pi) -- (-pi,0) -- cycle;
    \end{tikzpicture}
    \hfill
    \begin{tikzpicture}[scale=0.9]
        \begin{scope}[decoration={snake,amplitude=.4mm,segment length=2mm,post length=2mm},>=Latex,thick,transparency group,opacity=0.7,accent]
            \foreach \t in {-30,0,30}{                
                \draw[decorate,->] ($(0,{rad(\t)})$) -- ($({(pi-rad(\t))/2},{(pi+rad(\t))/2})+(-0.1,-0.1)$);
                \draw[decorate,->] ($(0,{rad(\t)})$) -- ($({-(pi-rad(\t))/2},{(pi+rad(\t))/2})+(0.1,-0.1)$);
            };
        \end{scope}

        \fill[color=accent!90!white,very thick] (0,-pi) .. controls ($(pi/2,-pi/4)$) and ($(0,3*pi/4)$) .. (0,pi) .. controls ($(0,3*pi/4)$) and ($(-pi/2,-pi/4)$) .. (0,-pi);
        
        \foreach \t in {-45, 50}{                
            \draw[smooth,accent!65!black,very thick] ($({(-pi+rad(\t))/2},{(pi+rad(\t))/2})$) -- ($(0,{rad(\t)})$) -- ($({(pi-rad(\t))/2},{(pi+rad(\t))/2})$); 
        };

        \draw[thick] (0,-pi) -- (pi,0) -- (0,pi) -- (-pi,0) -- cycle;
    \end{tikzpicture}
    \hfill\null
    \caption{Different definitions of mass in general relativity put observers in different places. The ADM notion of mass considers measurements of the mass at the spacelike infinity, as in the left picture. This construction leads to a conserved notion of mass, for the emission of gravitational waves (for example) is intercepted by future spatial slices. The Bondi definition, however, considers observers who can detect gravitational waves; asymptotically null hypersurfaces can avoid radiation emitted to infinity. Thus, the Bondi mass can decrease due to the emission of gravitational radiation, while the ADM mass is constant by construction. This figure is inspired by Fig. 14 in Ref. \onlinecite{penrose1968StructureSpacetime}.}
    \label{fig: concepts-of-mass}
\end{figure}

At first sight, it may seem reasonable to seek a definition that is conserved, as it would restore meaning to total energy conservation for spacetimes that lack a timelike isometry. On second thought, such a definition would prevent us from asserting that the observation of gravitational radiation carries energy away from the sources. This distinction is the reason why the Arnowitt--Deser--Misner (ADM) definition \cite{arnowitt1962DynamicsGeneralRelativity,*arnowitt2008RepublicationDynamicsGeneral} does not always coincide with Bondi's \cite{bondi1962GravitationalWavesGeneral}. The former places an observer at spatial infinity, leaving no room for energy fluxes to escape a spacelike surface enclosed by it, whereas the latter considers an observer who can detect gravitational waves and is therefore located at null infinity. This is illustrated in Fig. \ref{fig: concepts-of-mass}. Notice how this behavior is similar to electrodynamics: static electromagnetic fields decay slower as \(r\) goes to infinity than radiative fields. Observers that are spacelike separated from the sources cannot detect the faster-decaying radiative fields.

It is worth inspecting these two definitions more closely. The ADM mass is motivated by a Hamiltonian formulation of general relativity. By considering a suitable Cauchy surface in spacetime, one can take the induced metric on the surface and its ``time-derivative'' (more appropriately, the surface's extrinsic curvature) as canonical coordinates on a field-theoretic phase space. While diffeomorphism-invariance leads to Hamiltonian constraints that need to be dealt with \cite{arnowitt1962DynamicsGeneralRelativity,dirac2001LecturesQuantumMechanics}, it is possible to understand the canonical generators of symmetry transformations as conserved Noether charges. For instance, the ADM mass is precisely the Hamiltonian of general relativity after this construction is carried out.

It is more difficult to extend the same interpretation to the Bondi mass. As previously mentioned, gravitational radiation (and other sorts of radiation) can reach null infinity, and thus change the value of a would-be Noether charge. In this sense, one could say that physics at null infinity is similar to an open system, whereas physics at spatial infinity is a closed system. Although no new information can enter or leave spatial infinity due to causality constraints (and hence we can define a Hamiltonian in a more-or-less direct way), null infinity is often perturbed by the arrival (or departure) of radiation. In this more general setup, is it possible to understand mass in terms of a ``Noether-like charge'' associated with a symmetry at infinity? In other words, can we define the Bondi mass (and related quantities) in a Hamiltonian formalism?

This was addressed (in more generality) by \textcite{wald2000GeneralDefinitionConserved} by working with a phase space formulation of field theories. We first notice that, in classical mechanics with finitely many degrees of freedom, we know the Hamilton equations imply that the canonical coordinates \(q\), their conjugate momenta \(p\), and the Hamiltonian \(H\) are related by
\begin{subequations}
    \begin{align}
        \var{H} &= \dot{q} \var{p} - \dot{p} \var{q}, \\
        &= \Omega(\var{z}, \dot{z})|_z,
    \end{align}
\end{subequations}
where \(z = (q,p)\) is a point in phase space and \(\Omega\) denotes the symplectic form acting on the vector fields \(\var{z}\) and \(\dot{z}\). If, in field theory, we are given a symmetry generator \(\tensor{\xi}{^a}\), we demand that the variation of a Hamiltonian whose flow follows that symmetry obey
\begin{equation}\label{eq: var-H-xi-integral-omega}
    \var{H}_{\xi} = \Omega_{\Sigma}(\var\phi,\pounds_\xi \phi)|_{\phi} = \int_{\Sigma} \vb*{\omega}(\var{\phi},\pounds_{\xi}\phi)|_{\phi}.
\end{equation}
for a spacetime three-form \(\vb*{\omega}\), which can be obtained directly from the Lagrangian describing the theory (see, for example, Ref. \onlinecite{wald2000GeneralDefinitionConserved}). After imposing the equations of motion, one can show that \(\var{H}_{\xi}\) is given by an integral over a two-dimensional surface \(\partial\Sigma\) in spacetime.

We wish to define \(H_{\xi}\) itself, not only its variation. \(H_{\xi}\) could, in principle, also be given by an integral over \(\partial\Sigma\). A necessary condition for the existence of \(H_{\xi}\) is that 
\begin{equation}\label{eq: vanishing-symplectic-flux-boundary}
    0 = \int_{\partial\Sigma} \xi \vdot \vb*{\omega}(\var_1\phi,\var_2\phi)|_{\phi}, 
\end{equation}
where \(\tensor{(\xi \vdot\vb*{\omega})}{_a_b} = \tensor{\xi}{^c}\tensor{\omega}{_c_a_b}\) and \(\var_1\phi\) and \(\var_2\phi\) are two field variations satisfying the linearized equations of motion.

Equation \eqref{eq: vanishing-symplectic-flux-boundary} states that the ``symplectic flux'' through \(\partial\Sigma\) vanishes. In other words, the physical system is ``sufficiently closed'' for us to define a Noether-like charge. Recall that $\vb*{\omega}$, built from a Lagrangian, only carries information about the dynamics, whereas Eq.~\eqref{eq: vanishing-symplectic-flux-boundary} directly refers to the vector \(\tensor{\xi}{^a}\). Hence, even if $\vb*{\omega}$ fails to decay sufficiently fast at infinity, the symplectic flux condition~\eqref{eq: vanishing-symplectic-flux-boundary} can still be satisfied if $\partial \Sigma$ can be chosen tangent to the symmetry generator $\xi^a$.

We are now ready to revisit the problem of defining mass in asymptotically flat spacetimes. We consider our phase space to be comprised of asymptotically flat metrics in a suitable sense and equip it with the three-form \(\vb*{\omega}\) obtained from the Einstein--Hilbert action. There are two geometric regions of interest. The first is spatial infinity \(i^0\), a point spacelike separated to the entirety of the spacetime on which the ADM mass is defined. The second is future\footnote{Of course, we could just as well choose to work with past null infinity with minimal modifications.} null infinity \(\mathscr{I}^+ \simeq \mathbb{R} \times \mathbb{S}^2\), with the Bondi mass defined on the cross sections of \(\mathscr{I}^+\). The key mathematical distinction between the ADM and Bondi masses arises at this point. \(\vb*{\omega}\) vanishes identically at \(i^0\) because there are no degrees of freedom at spatial infinity. Radiation cannot reach \(i^0\), so the gravitational field is static and completely defined by the matter distribution. As a consequence, every symmetry generator \(\tensor{\xi}{^a}\) defines a conserved quantity. For asymptotic time translations, that quantity is the ADM mass \(H_{\text{ADM}}\). By considering different Cauchy surfaces with the same boundary (a topological sphere tending to \(i^0\)), the result will always be the same, owing to the absence of a symplectic flux through the boundary. Hence, the ADM mass is conserved.

This is in contrast with the Bondi mass. At null infinity, \(\vb*{\omega}\) no longer vanishes identically and we get symplectic fluxes through \(\mathscr{I}^+\) due to the occurrence of radiation. The Bondi mass consequently changes as one moves along different spherical cross sections of \(\mathscr{I}^+\). We no longer have conservation of mass because radiation can carry energy away through null infinity.

The integrals over \(\partial\Sigma\) that define the ADM or Bondi masses are strikingly similar to Eq. \eqref{eq: active-mass-newton}, which defines the (Newtonian) mass nonlocally in terms of an integral at infinity. However, this expression can be ``localized''. The acceleration caused by a mass distribution on a test particle is given by 
\begin{equation}
    \div\vb{a} = - 4 \pi G \rho, 
\end{equation}
where \(\rho\) is the mass distribution of the source. Using Stokes' theorem in Eq. \eqref{eq: active-mass-newton} thus leads to 
\begin{equation}
    M = \int_{\mathbb{R}^3} \rho \dd{V},
\end{equation}
with \(\dd{V}\) the volume element on \(\mathbb{R}^3\). In this sense, the mass \(M\) is the sum of its parts---the total mass of an object is obtained by summing over the masses of its components. 

It would be interesting to perform a similar ``localization procedure'' in general relativity. We can use Stokes' theorem again to write some charge \(H_{\xi}\) as an integral over \(\Sigma\) and identify the integrand as a ``mass density.'' This would allow us to interpret the pictures in Fig. \ref{fig: concepts-of-mass} more faithfully, with the ADM and Bondi masses given as volume integrals over three-dimensional hypersurfaces.

This procedure is not as straightforward as it may seem. Consider, for example, the extended Schwarzschild geometry, which has a nonzero mass despite being a vacuum solution. A naive application of Stokes' theorem would state that the ``mass density'' should be identically zero, but the total mass is positive. How is this possible? 

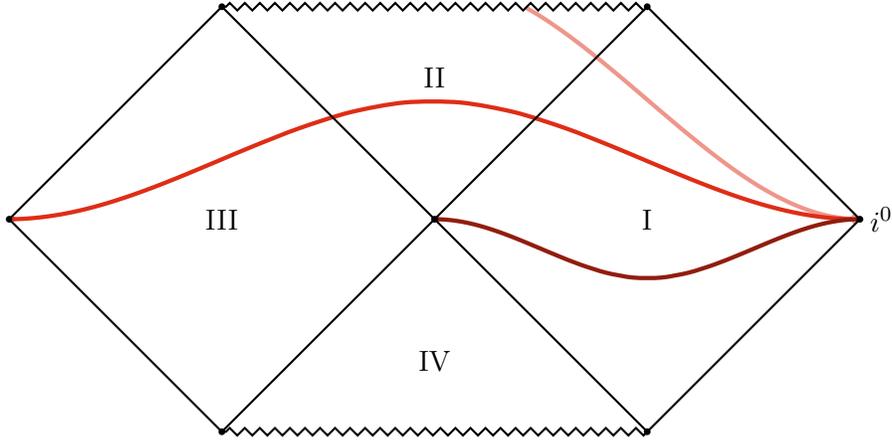
\begin{figure}
    \centering
    \begin{tikzpicture}[scale=0.9]
        \begin{scope}[decoration = {zigzag,segment length = 2mm, amplitude = 0.5mm}]
            \clip (0,pi) decorate{-- (-2*pi,pi)} -- (0,-pi) -- (pi,0) -- cycle;
            \draw[domain=0:50,smooth,variable=\r,samples=400,accent!50!white,ultra thick] plot ({2*rad(atan(tan(50/2)+\r))-2*rad(atan(tan(50/2)-\r))-pi},{4.05*rad(atan(tan(50/2)+\r))+4.05*rad(atan(tan(50/2)-\r))});
        \end{scope}

        \draw[domain=-50:50,smooth,variable=\r,samples=400,accent,ultra thick] plot ({2*rad(atan(tan(50/2)+\r))-2*rad(atan(tan(50/2)-\r))-1.01*pi},{2*rad(atan(tan(50/2)+\r))+2*rad(atan(tan(50/2)-\r))});
        \draw[domain=-50:50,smooth,variable=\r,samples=400,accent!65!black,ultra thick] plot ({rad(atan(tan(-50/2)+\r))-rad(atan(tan(-50/2)-\r))},{rad(atan(tan(-50/2)+\r))+rad(atan(tan(-50/2)-\r))});
        
        \draw[decoration = {zigzag,segment length = 2mm, amplitude = 0.5mm},thick,decorate] (0,pi) -- (-2*pi,pi);
        
        \draw[decoration = {zigzag,segment length = 2mm, amplitude = 0.5mm},thick,decorate] (0,-pi) -- (-2*pi,-pi);

        \draw[thick] (0,-pi) -- (pi,0) -- (0,pi) -- (-pi,0) -- cycle;

        \draw[thick] (-2*pi,-pi) -- (-pi,0) -- (-2*pi,pi) -- (-3*pi,0) -- cycle;

        \fill (-2*pi,pi) circle (0.05);
        \fill (0,pi) circle (0.05);

        \fill (-3*pi,0) circle (0.05);
        \fill (-pi,0) circle (0.05);
        \fill (pi,0) circle (0.05);

        \fill (-2*pi,-pi) circle (0.05);
        \fill (0,-pi) circle (0.05);

        \node at (0,0) {I};
        \node at (-pi,2*pi/3) {II};
        \node at (-2*pi,0) {III};
        \node at (-pi,-2*pi/3) {IV};

        \node[anchor=west] at (pi,0) {\(i^0\)};
    \end{tikzpicture}
    \caption{Different hypersurfaces on which one could evaluate the ADM mass as a hypersurface integral in the maximal analytic extension of Schwarzschild spacetime. While one wants to convert the integral over spatial infinity in Region I (indicated as \(i^0\)), all hypersurfaces require a boundary term once Stokes' theorem is applied. The lightest-colored hypersurface ends at a small surface before the singularity. In the second hypersurface (middle-dark), there is a nontrivial contribution due to spatial infinity in Region III. Finally, in the darkest hypersurface, the boundary contribution is due to the bifurcation surface.}
    \label{fig: schwarzschild}
\end{figure}

One is tempted to establish a parallel with pointlike particles in flat-space classical electrodynamics. There, too, one has \(\div\vb{E}=\rho_\text{electric}/\epsilon_0=0\) everywhere, except at the origin. The total charge is not zero because the integrand contains a pole in any integration three-surface. Thus, the best one can do when applying Stokes' theorem is to trade the sphere at infinity for an arbitrarily small closed surface surrounding the origin. The three-dimensional region between this small surface and infinity does not contribute because the integrand $\div\vb E$ is zero on it. It is reasonable then to interpret that the whole charge is located at the singular point. However, the issue is subtler in a curved spacetime because spacelike hypersurfaces can bypass the singularity\footnote{We could, in principle, work with nonspacelike surfaces as well, but it is still impossible to convert the integral into a strictly volumetric integral.}. Consider, for example, the illustration in Fig. \ref{fig: schwarzschild}. We want to consider the ADM mass associated with Region I (this coincides with the Komar and Bondi masses in Region I for this geometry). Since we must not choose to integrate over a hypersurface extending up to the singularity (which is not a part of spacetime), much like in the flat-space electrostatic analogy, we are forced to consider a boundary ``arbitrarily close to the singularity.'' Again, this boundary originates the nonvanishing contribution to the mass. If we extend the hypersurface into Region III, the nonzero contribution emanates from Region III's spatial infinity. A third alternative is to bring the boundary to the bifurcation surface. In all cases, the presence of boundary terms forces us to admit nonvanishing masses, despite the spacetime being a vacuum solution at all points.

This simple geometry illustrates that mass is more than the sum of its parts in relativity. Not only does the total mass contain local matter contributions in spacetime, but it may also receive a nonlocal contribution from the geometry of spacetime itself. The astute reader will also notice that the total electric charge in a curved spacetime behaves similarly. In the extended Reissner--Nordström geometry, the total charge can be similarly calculated as an integral on a two-surface $\partial\Sigma$ at infinity. After applying Stokes' theorem, the resulting integrand $\dd{\!\star\!\vb{F}}$, where $\vb{F}$ is the Faraday tensor, vanishes identically, and no spacelike three-surface whose boundary is $\partial\Sigma$ contains a singularity. The charge is nowhere to be localized!

There are, surely, spacetimes in which these difficulties do not arise. For example, a spacetime describing an isolated stable star, with no horizons or singularities. However, even in these cases, the integrand does not coincide with the local energy density measured by any observer. This is anticipated: if a system is bound by gravity, the total mass of the system should be smaller than the sum of the masses of its constituents. Startlingly, mass can also be less than the sum of its parts.

With that in mind, we wish to quantify the gravitational binding energy. Could it be so negative that the total mass of a relativistic system becomes negative as well? There are two possible scenarios of interest. In the first case, we assume that all possible matter constituents have a positive mass, meaning that their acceleration points in the same direction as the forces acting on them. In this case, the gravitational binding energy could be identified as the culprit for the negative mass. In the second case, we consider matter constituents with negative mass, which can further contribute to the violation of the positivity of energy. Can a macroscopic mass actually become negative in either scenario?

To formulate the question more precisely, we must rigorously define ``positive mass components.'' In relativity, this comes in the form of energy conditions. More accurately, the weak energy condition (WEC) corresponds to the assumption that all observers measure non-negative mass-energy densities at all points in spacetime. Under some additional global hypotheses, \textcite{witten1981NewProofPositive} showed that the WEC forbids macroscopic negative masses. The proof reveals a remarkable feature: the total mass can be written as
\begin{equation}
M=\frac{1}{16\pi G}\int_{\Sigma} \left\{h^{ab}(\nabla_a\epsilon)^\dagger(\nabla_b\epsilon)+G_{cd}t^c\epsilon^\dagger\gamma^dt_e\gamma^e\epsilon\right\}\dd{\mathcal{V}},
    \label{eq: witten}
\end{equation}
where \(G_{cd}\) is the Einstein tensor, \(h_{ab}\) is the induced metric on \(\Sigma\), \(\dd{\mathcal{V}}\) is its volume element, and \(\epsilon\) is a spinor field satisfying the equation \(i h^{ab}\tensor{\gamma}{_a}\tensor{\nabla}{_b}\epsilon=0\)\footnote{Here, $\gamma^a\equiv e_\mu^a\gamma^\mu$, where $e_\mu^a$ are frame fields and $\gamma^\mu$ are the standard Dirac gamma matrices in tangent space.}. The first term in Eq. \eqref{eq: witten}, independent of the energy-momentum tensor, turns out to be nonnegative. This observation highlights the difficulty of localizing mass in general relativity. While textbook arguments (see, \emph{e.g.}, Ref. \onlinecite{wald1984GeneralRelativity}) show that the gravitational binding energy of a spherically symmetric configuration is negative, as expected, the Ricci-independent part of the geometry contributes positively to the mass. Since the propagating degrees of freedom of the gravitational field lie in the Ricci-independent part of the geometry, we can say that the gravitational degrees of freedom contribute positively to the mass\footnote{The positive mass in the maximally extended Schwarzschild geometry we discussed above makes this feature not wholly surprising.}.

As with any theorem, though, Witten's is only as good as its hypotheses. Notably, the weak energy condition is not verified in quantum theory, which raises the question of whether the conclusion could be reversed in that setting. It is desirable to obtain a more general positivity result that is valid under conditions that are more friendly toward quantum mechanics.

One may be tempted to think that the positive mass theorem is put in a similar position to the singularity theorems, which also assume energy conditions. Many singularity theorems have been adapted to a quantum-friendly setup after their hypotheses were modified. The positive mass theorem is more challenging. For the singularity theorems, the weak energy condition can be relaxed in favor of its ``averaged version,'' stipulating that only the integral of the energy density along a complete timelike curve is nonnegative. Strong arguments support that these averages are verified under mild hypotheses in quantum field theory \cite{fewster2006AveragedEnergyInequalities, kontou2020EnergyConditionsGeneral}. For an analogous generalization of Witten's theorem, one would need similar averages not along timelike (or null) curves but rather over three-dimensional spacelike surfaces (see Eq. (46) of Ref. \onlinecite{witten1981NewProofPositive}). Such averages, however, can still be violated in quantum field theory \cite{ford2002SpatiallyAveragedQuantum}. An ingenious argument by Penrose, Sorkin, and Woolgar (PSW) in Ref. \onlinecite{penrose1993PositiveMassTheorem} overcame this difficulty. We now summarize their findings.

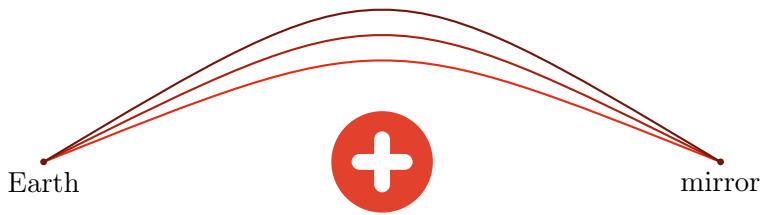
\begin{figure}
    \centering
    \begin{tikzpicture}[scale=0.9]
        \fill[accent!90!white] (0,0) circle (0.75);
        \draw[line width=6pt,{Round Cap}-{Round Cap},white] (-0.45,0) -- (0.45,0);
        \draw[line width=6pt,{Round Cap}-{Round Cap},white] (0,-0.45) -- (0,0.45);

        \draw[thick,accent] (-5,0) .. controls (0,2) .. (5,0);
        \draw[thick,accent!75!black] (-5,0) .. controls (0,2.5) .. (5,0);
        \draw[thick,accent!50!black] (-5,0) .. controls (0,3) .. (5,0);

        \fill[accent!50!black] (-5,0) circle (0.05);
        \node[anchor=north] at (-5,0) {Earth};
        \fill[accent!50!black] (5,0) circle (0.05);
        \node[anchor=north] at (5,0) {mirror};
    \end{tikzpicture}
    \caption{Diagram illustrating the Shapiro time delay effect \cite{shapiro1964FourthTestGeneral}. A laser is sent from the Earth to a mirror and follows the same path backwards. The coordinate time (as measured on Earth) for the full trajectory is smaller for light rays that pass farther away from the Sun. In this picture, the darker lines (illustrating geodesics farther away from the mass) are measured to be faster with respect to coordinate time when making a round-trip.}
    \label{fig: shapiro}
\end{figure}

Within relativity, the presence of a mass can affect clocks in its vicinity. Notably, this can lead to time dilation effects, such as the Shapiro time delay \cite{shapiro1964FourthTestGeneral}. This effect predicts that null geodesics coming closer to a mass run slower (in coordinate time) than geodesics that travel farther away from it. See pages 146--148 of Ref. \onlinecite{wald1984GeneralRelativity}, for example, and the illustration on Fig. \ref{fig: shapiro}. 

The first key observation, due to PSW, is the effect of reversing the sign of the mass in the standard Shapiro time-delay formula. For a negative mass, the fastest null geodesics must pass very close to the mass. In particular, they travel through the bulk of the spacetime rather than being pushed to its conformal boundary. A negative total mass in spacetime thereby requires that the fastest null geodesic run through the bulk. If one forbids them from entering spacetime, one also forbids negative masses.

The next key observation is the known theorem stating that the fastest null geodesic between two surfaces must always be orthogonal to the surfaces and achronal, meaning that no two of its points can be connected by a timelike curve. Since achronal curves do not contain a pair of conjugate points \cite{hawking1973LargeScaleStructure}, if one can establish that conjugate points are present in all null geodesics in the bulk, it will follow that the fastest null curve cannot run through the bulk, and hence there cannot be a negative mass. 

After combining these ingredients, the proof techniques for the new positive mass theorem due to PSW resemble those of the singularity theorems. PSW invoked a result due to \textcite{borde1987GeodesicFocusingEnergy} that the averaged null energy condition (when complemented by the null generic condition) is sufficient to enforce the occurrence of conjugate points to close the argument. Moreover, since achronal geodesics are the only candidates for the fastest causal curves, the energy condition can be simplified to the achronal averaged null energy condition, which is believed to hold in quantum field theory \cite{kontou2020EnergyConditionsGeneral,wall2010ProvingAchronalAveraged}.

With some poetic license, the Penrose--Sorkin--Woolgar theorem states that one cannot create macroscopic negative masses in quantum field theory. This contrasts with how easily local negative energy densities are constructed in the same framework \cite{casimir1948AttractionTwoPerfectly,fewster2012LecturesQuantumEnergy,costa2022CanQuantumMechanics}. Yet, the PSW theorem still requires ``well-behaved matter'' (\emph{i.e.}, an energy condition) to reach its conclusion. One may thus still inquire whether general relativity prohibits negative masses by itself, without appealing to energy conditions.  We finally turn our attention to the second scenario. 

We partially addressed this question in collaboration with Landulfo~\cite{aguiaralves2025PositiveMassGeneral}. We examined previously considered \cite{novikov2018StarsCreatingGravitational} spherically symmetric stellar configurations that exhibit examples of non-singular spacetimes with negative masses. We studied whether such configurations are stable using standard techniques from stellar stability analysis \cite{chandrasekhar1964DynamicalInstabilityApJ,*chandrasekhar1964DynamicalInstabilityApJErratum,chandrasekhar1964DynamicalInstabilityPRL,*chandrasekhar1964DynamicalInstabilityPRLErratum,bardeen1966CatalogueMethodsStudying}, and showed that barotropic negative-mass stars are generically unstable in general relativity. The result does not assume the validity of any energy conditions, and, in this sense, suggests that general relativity abhors negative masses. Even if a macroscopic negative mass configuration were to be created, it would be destined to lose its structure in a finite time. For a solar-sized configuration, the instability timescale is about three minutes \cite{aguiaralves2025PositiveMassGeneral}. While larger structures have larger instability timescales, they raise the question of how they could have been created in the first place.

At first, we thought that mass could be simply the sum of its parts, but we now realize that it can also be more or less than that. But even when it is less than the sum, it is resolute not to be negative. Both Data and humans deserve to be appreciated as a whole. Their idiosyncrasies are emergent phenomena, and so are the whims of masses in general relativity. To quantify the whole is to boldly measure what no local observer has measured before!

\begin{acknowledgments}
    B. A. C. acknowledges R. Adhikari for checking his calculations. The work of N. A. A. was supported by the Coordenação de Aperfeiçoamento de Pessoal de Nível Superior---Brasil (CAPES)---Finance Code 001.
\end{acknowledgments}

\bibliography{bibliography} 
\end{document}